\documentclass[oneside, sn-nature, numbered, twocolumn]{sn-jnl}

\usepackage{physics}
\usepackage[english]{babel}
\usepackage[utf8]{inputenc}
\usepackage{tabularray}
\usepackage{longtable}
\usepackage{indentfirst}
\usepackage{graphicx}
\usepackage{multirow}
\usepackage{amsmath,amssymb,amsfonts}
\usepackage{amsthm}
\usepackage{mathrsfs}
\usepackage[title]{appendix}
\usepackage{xcolor}
\usepackage{textcomp}
\usepackage{manyfoot}
\usepackage{booktabs}
\usepackage{algorithm}
\usepackage{algorithmicx}
\usepackage{algpseudocode}
\usepackage{listings}
\usepackage{siunitx}
\usepackage{float}
\usepackage{nomencl}
\makenomenclature
\setcitestyle{numbers}
\usepackage{adjustbox}
\usepackage[labelsep=period]{caption}
\usepackage{comment}
\usepackage{booktabs}

\usepackage{geometry}
\begin{document}
\title[Article Title]{Sustainability-Driven Exploration of Topological Materials}

\author[1,2]{\fnm{Artittaya} \sur{Boonkird}}\email{tiyab00@mit.edu}
\author[1,3]{\fnm{Nathan C} \sur{Drucker}}
\author[1,2]{\fnm{Manasi} \sur{Mandal}}
\author[1,2]{\fnm{Thanh} \sur{Nguyen}}
\author[4]{\fnm{Jingjie} \sur{Yeo}}
\author[5]{\fnm{Vsevolod} \sur{Belosevich}}
\author[6]{\fnm{Ellan} \sur{Spero}}
\author[6]{\fnm{Christine} \sur{Ortiz}}
\author[5]{\fnm{Qiong} \sur{Ma}}
\author[7]{\fnm{Liang} \sur{Fu}}
\author[8]{\fnm{Tomas} \sur{Palacios}}
\author[1,2]{\fnm{Mingda} \sur{Li}}\email{mingda@mit.edu}

\affil[1]{\orgdiv{Quantum Measurement Group}, \orgname{MIT}, \orgaddress{\city{Cambridge}, \state{MA} \postcode{02139}, \country{USA}}}
\affil[2]{\orgdiv{Department of Nuclear Science and Engineering}, \orgname{MIT}, \orgaddress{\city{Cambridge}, \state{MA} \postcode{02139}, \country{USA}}}
\affil[3]{\orgdiv{John A. Paulson School of Engineering and Applied Sciences}, \orgname{Harvard University}, \orgaddress{\city{Cambridge}, \state{MA} \postcode{02138}, \country{USA}}}
\affil[4]{\orgdiv{Sibley School of Mechanical and Aerospace Engineering}, \orgname{Cornell University}, \orgaddress{\city{Ithaca}, \state{NY} \postcode{14850}, \country{USA}}}
\affil[5]{\orgdiv{Department of Physics}, \orgname{Boston College}, \orgaddress{\city{Chestnut Hill}, \state{MA} \postcode{02467}, \country{USA}}}
\affil[6]{\orgdiv{Department of Materials Science and Engineering}, \orgname{MIT}, \orgaddress{\city{Cambridge}, \state{MA} \postcode{02139}, \country{USA}}}
\affil[7]{\orgdiv{Department of Physics}, \orgname{MIT}, \orgaddress{\city{Cambridge}, \state{MA} \postcode{02139}, \country{USA}}}
\affil[8]{\orgdiv{Department of Electrical Engineering and Computer Science}, \orgname{MIT}, \orgaddress{\city{Cambridge}, \state{MA} \postcode{02139}, \country{USA}}}

\abstract{Topological materials are at the forefront of quantum materials research, offering tremendous potential for next-generation energy and information devices. However, current investigation of these materials remains largely focused on performance and often neglects the crucial aspect of sustainability. Recognizing the pivotal role of sustainability in addressing global pollution, carbon emissions, resource conservation, and ethical labor practices, we present a comprehensive evaluation of topological materials based on their sustainability and environmental impact. Our approach involves a hierarchical analysis encompassing cost, toxicity, energy demands, environmental impact, social implications, and resilience to imports. By applying this framework to over 16,000 topological materials, we establish a sustainable topological materials database. Our endeavor unveils environmental-friendly topological materials candidates which have been previously overlooked, providing insights into their environmental ramifications and feasibility for industrial scalability. The work represents a critical step toward industrial adoption of topological materials, offering the potential for significant technological advancements and broader societal benefits.}

\maketitle

Advanced functional and quantum materials are at the foundation of improved technological performance. For microelectronics, the introduction of high-$\kappa$ dielectrics mitigates leakage current for CMOS technology and plays a pivotal role in sustaining Moore’s law \cite{wilk2001}. For optoelectronics, blue light-emitting diodes (LEDs) have revolutionized lighting by offering superior efficiency and longevity \cite{akasaki2015,nakamura2015} over prior lighting solutions. This breakthrough is of paramount significance as lighting accounts for 20\% of the global electricity consumption \cite{UN2017}. For magnetic devices, the utilization of giant magnetoresistance (GMR) has become ubiquitous for magnetic sensing applications, exemplified by its role in hard disk drives \cite{fert2008,grunberg2008}, while magnetic tunneling junctions (MTJ) serve as the foundation for magnetoresistive random-access memory (MRAM). For clean energy harvesting, multi-junction solar cells and emerging photovoltaic technologies like perovskites are rapidly advancing toward heightened efficiency \cite{geisz2020,kim2020,NREL}. 
Nonetheless, significant technological challenges persist. From an energy standpoint, around 70\% of generated energy ultimately dissipates as waste heat. From an information standpoint, the escalating energy demand for computational resources is projected to surpass the entire global energy production in decades \cite{semi1}. Therefore, there is a constant need to further discover, design, and optimize next-generation functional materials for future energy and information applications.

\begin{figure*}
 \centering
 \includegraphics[width=1.7\columnwidth]{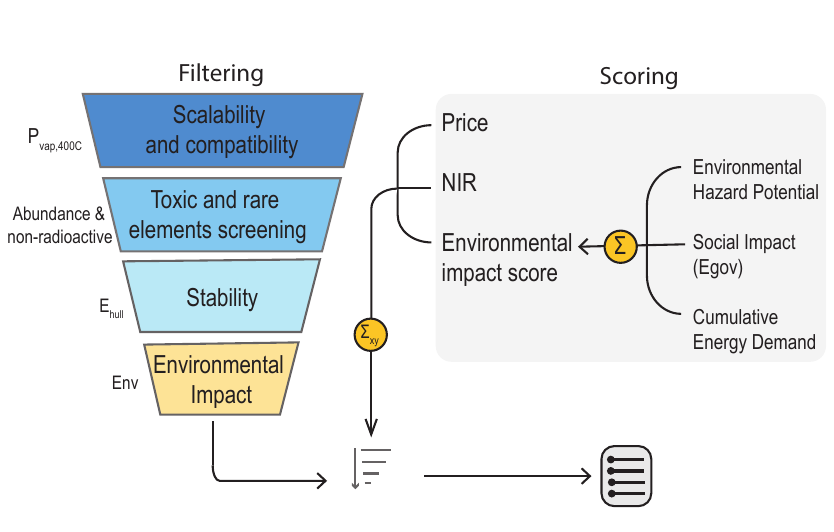}
 \caption{\textbf{Down-selection and scoring of topological materials.} The selection is limited to stable materials ($E_{\text{hull}}<0$) characterized by non-toxic elements, a modest melting temperature $T_{\text{melt}}<2000\ \text{K}$, and low vapor pressure ($P_{\text{vap},\ 400^\circ \text{C}}$). These materials can then be sorted based on price, net import resilience (NIR), and environmental impact score (Env). When applied to the topological materials database, this process yields roughly 200 materials candidates from a pool of over 16,000.}
 \label{figure1}
\end{figure*}

One path to address these challenges is by improving and developing topological materials, a major category of quantum materials where the electronic states are robustly protected by topology \cite{hasan2010,armitage2018}. The discovery of topological materials has revolutionized our understanding of condensed matter physics and material science, as recognized by the 2016 Nobel Prize in Physics \cite{haldane2017}. Topological materials manifest many highly promising effects which have demonstrated potential to span a diverse spectrum of devices. These include but are not limited to ballistic electron transport without energy dissipation \cite{konig2007,chang2013} and ultrahigh electrical conductivity \cite{han2023} for microelectronics, a topology-inspired laser \cite{harari2018,bandres2018} and non-reciprocal light emission \cite{zhao2020,tsurimaki2020} for optoelectronics, low-power current-induced switching for spintronics \cite{yang2021,han2023_spin}, broadband solar energy harvesting \cite{sodemann2015,ma2019}, high-power-factor waste heat recovery \cite{han2020} and energy storage \cite{luo2022}, among others. 

Two principal reasons make topological materials particularly attractive for next-generation energy and information applications. One, many topological-related effects are shown to exist at room temperature \cite{kumar2021,shumiya2022} since the topological protection can be robust enough to prevail over thermal fluctuations. Two, there are plenty of topological materials candidates available. Efficient topology screening methods such as topological quantum chemistry have identified more than 20,000 topological materials, with even more materials candidates through high-throughput calculations \cite{Database}. However, many mainstream topological materials are largely or solely driven by performance indicators, with substantially less attention paid to sustainability or industrial feasibility (see Table \ref{tab2} for a few prototypical topological materials). In light of this, the importance of carefully selecting topological materials with considerations of environmental sustainability and feasibility cannot be overemphasized \cite{titirici2022,epa2015}. Early consideration of environmental impact before massive adoption will help prevent previous chemical-related industrial mishaps such as CFCs (chlorofluorocarbons), DDT (dichlorodiphenyltrichloroethane), PFASs (per- and polyfluoroalkyl substances) from repeating. Moreover, an attentive eye toward sustainability would play a pivotal role in supporting the circular economy with abundant resources, longer durability, improved recyclability, and reduced-waste manufacturing.

In this work, we consider sustainability, potential for commercialization, and compatibility with currently available fabrication processes for the development of topological materials. We screen and score the topological materials by price, economic stability of the supply chain, and environmental impact of raw materials production. We identify about 200 candidates out of over 16,000 topological materials, many of which contain Earth-abundant elements such as Fe, Si, and Ni, that have a feasible prospect of being adopted by industry. In addition to identifying these sustainable topological materials, our methodology establishes a means to evaluate the sustainability of other quantum and advanced functional materials during early research and development stages.

\begin{figure*}
 \centering
 \includegraphics[width=2.0\columnwidth]{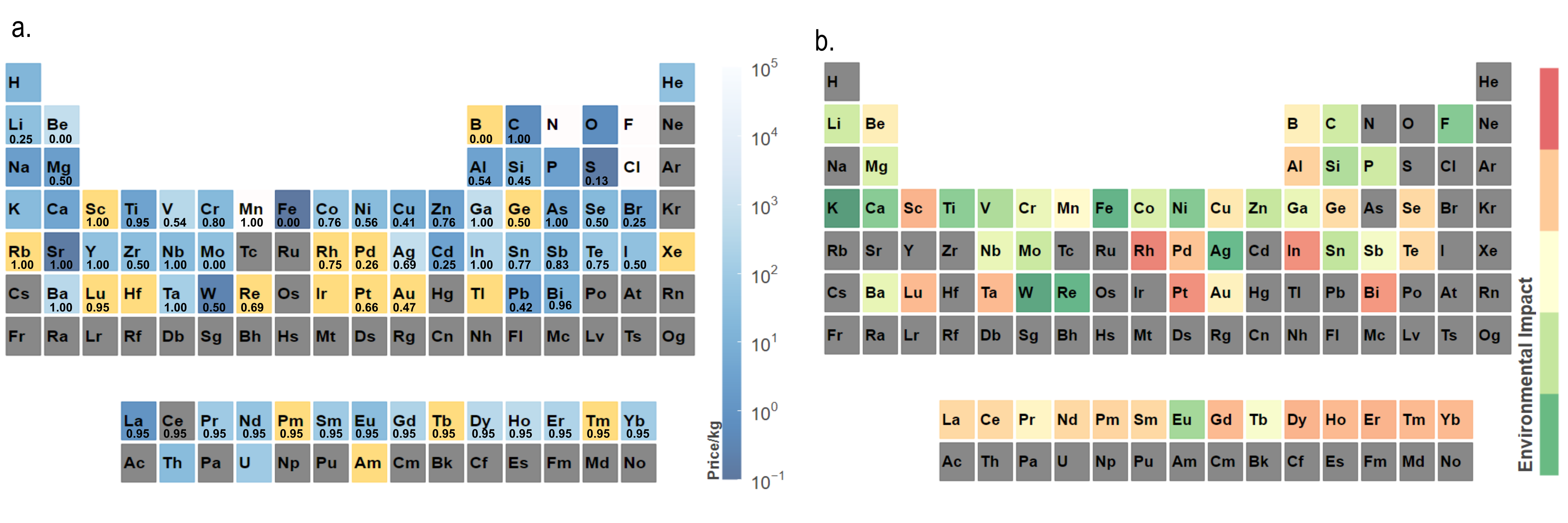}
 \caption{\textbf{\textbf{Elemental price and sustainability scores.}} \textbf{(a)} Elemental prices are represented in USD (in 2023) per kilogram for each element. Elements with unavailable price data are shaded in grey, while elements with prices beyond the scale are highlighted in yellow due to their prohibitively high cost. \textbf{(b)} Environmental impact scores per kilogram for each element on the periodic table, as applicable. Grey shading indicates the unavailability of environmental impact data. Abundant elements such as Fe, Ni, and Si are both relatively cheap and environmentally suitable.}
 \label{figure2}
\end{figure*}

\section*{Results}\label{sec2}
\begin{figure*}
 \centering \includegraphics[width=1.8\columnwidth]{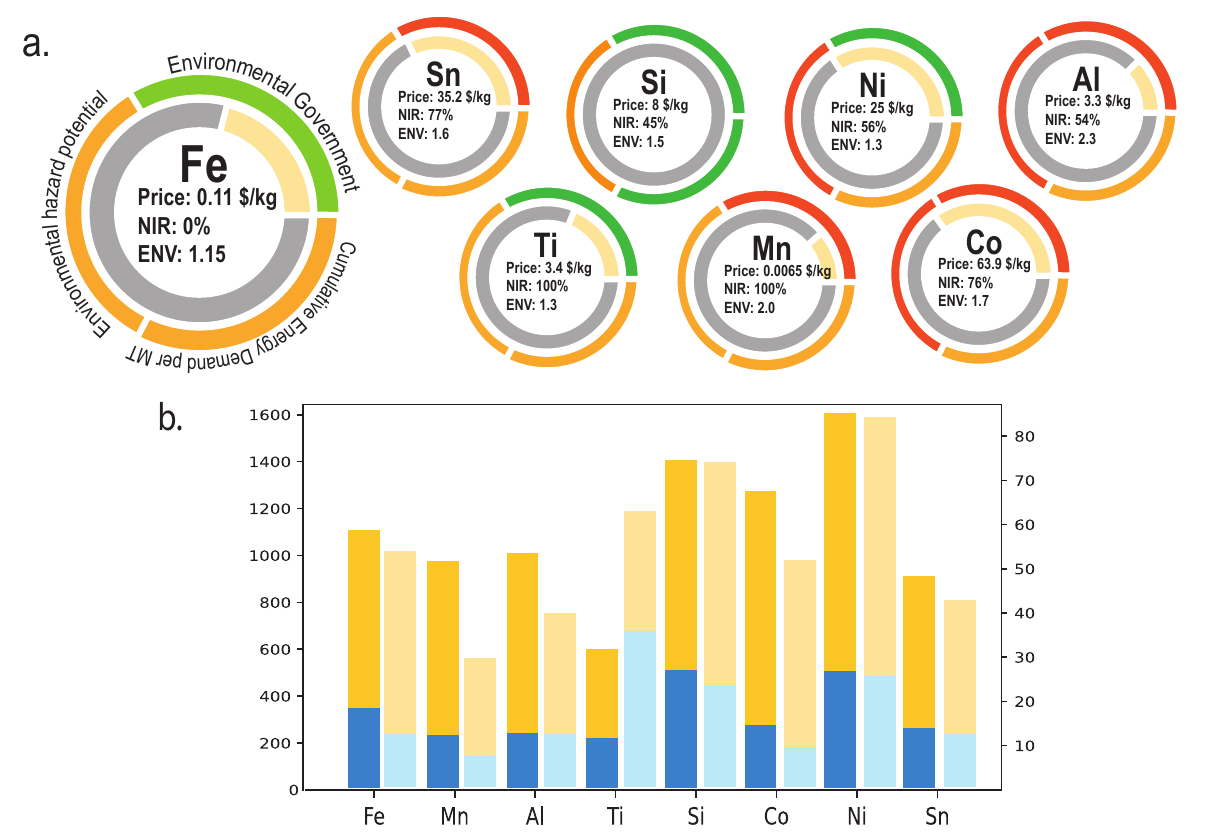} \caption{\textbf{\textbf{Sustainability analysis for elements most commonly found in topological materials.}} \textbf{(a)} Price, NIR, Env, environmental hazard potential, EGov score, cumulative energy demand per MT of pure elements (outer ring), and the recycling rate of the pure elements visualized as the proportion of yellow to grey in the inner ring. \textbf{(b)} The number of topological insulators (TI) and semimetal (SM) in the topological materials database before (left axis, darker colors) and after screening (right axis, lighter colors).}
 \label{figure3}
\end{figure*}

Materials from the theoretically-predicted, comprehensive topological materials database \cite{Database} are initially narrowed down by considering toxicity, rarity, and stability. The stability of elements is based on their calculated energy above the convex hull, and those with ($E_{hull} >$ 0 eV) \cite{bartel2022review}, are omitted. However, recognizing that stability alone does not guarantee large-scale producibility, a lower limit on synthesis temperature is defined based on raw material melting temperatures. Consequently, materials comprising elements with melting temperatures exceeding 2000 K are excluded from consideration.

These materials are further filtered based on considerations of their compatibility with existing production technologies. This criterion is especially important for materials with applications in the electronics industry, which has stringent manufacturing requirements. One criterion is the chemical stability of the material, ensuring that it does not react with the various processing chemicals employed in fabrication. Compounds containing alkali or alkaline earth metals, such as magnesium (Mg) and barium (Ba), are generally avoided due to their reactivity, as they can lead to contamination of the fabrication tools. Materials with high vapor pressure or metals that evaporate at the working temperature and pressure of CMOS production lines are prohibited to prevent contamination. This criterion applies to materials with vapor pressures exceeding 0.01 millitorr at 400$^\circ$C. However, determining the vapor pressure of materials at 400$^\circ$C requires experimental data. As a data-driven screening approach, materials composed of metals with vapor pressures above 0.01 millitorr are eliminated \cite{alcock1984vapour}. 

The final screening criterion is the environmental impact score (Env), which reflects environmental and social risks during the material production process. This assessment includes four factors: the environmental hazard potential (EHP), specific cumulative energy demand (CED), the environmental government index (EGov), and recyclability (R). This score is adapted from the ÖkoRess project report \cite{EnvCri}, providing an assessment of the environmental and social impact of raw materials production reflected by the potential hazard of mining, energy demand for extraction, and government performance in dealing with environmental problems scaled by the recycling rate of each pure element.

After the screening process, the materials are separately scored by price, net import resilience (NIR), and Env to obtain a list of topological materials for further study that have the potential to be industrialized based on these considerations. Including NIR and price reflects information related to feasibility of wide-scale development. The prices of both critical and non-critical materials are based on data from the U.S. Geological Survey \cite{ScienceBase_2017} and relevant commercial websites \cite{LondonMetalPrice,NonCriticalPrice, GasPrice}. The NIR of raw materials provides insight into the vulnerability of supply chain to disruptions. NIR reflects the extent to which a nation relies on the import of raw materials to meet its consumption demand. This value is derived by calculating the ratio of net imports of raw materials to apparent consumption and is reported as a percentage. In addition to scoring price and net import resilience, we assign the environmental impact as a score from 1 to 3 to each element, based on the aforementioned Env criteria. The pricing, NIR, and Env values for the individual elements on the periodic table are presented in Figure \ref{figure2}. Subsequently, for each topological material within the database, the corresponding price, NIR, and Env scores are determined through the application of weighted averages based on the elemental composition per kilogram of the material.

\begin{figure}
 \centering
 \includegraphics[width=1.20\columnwidth]{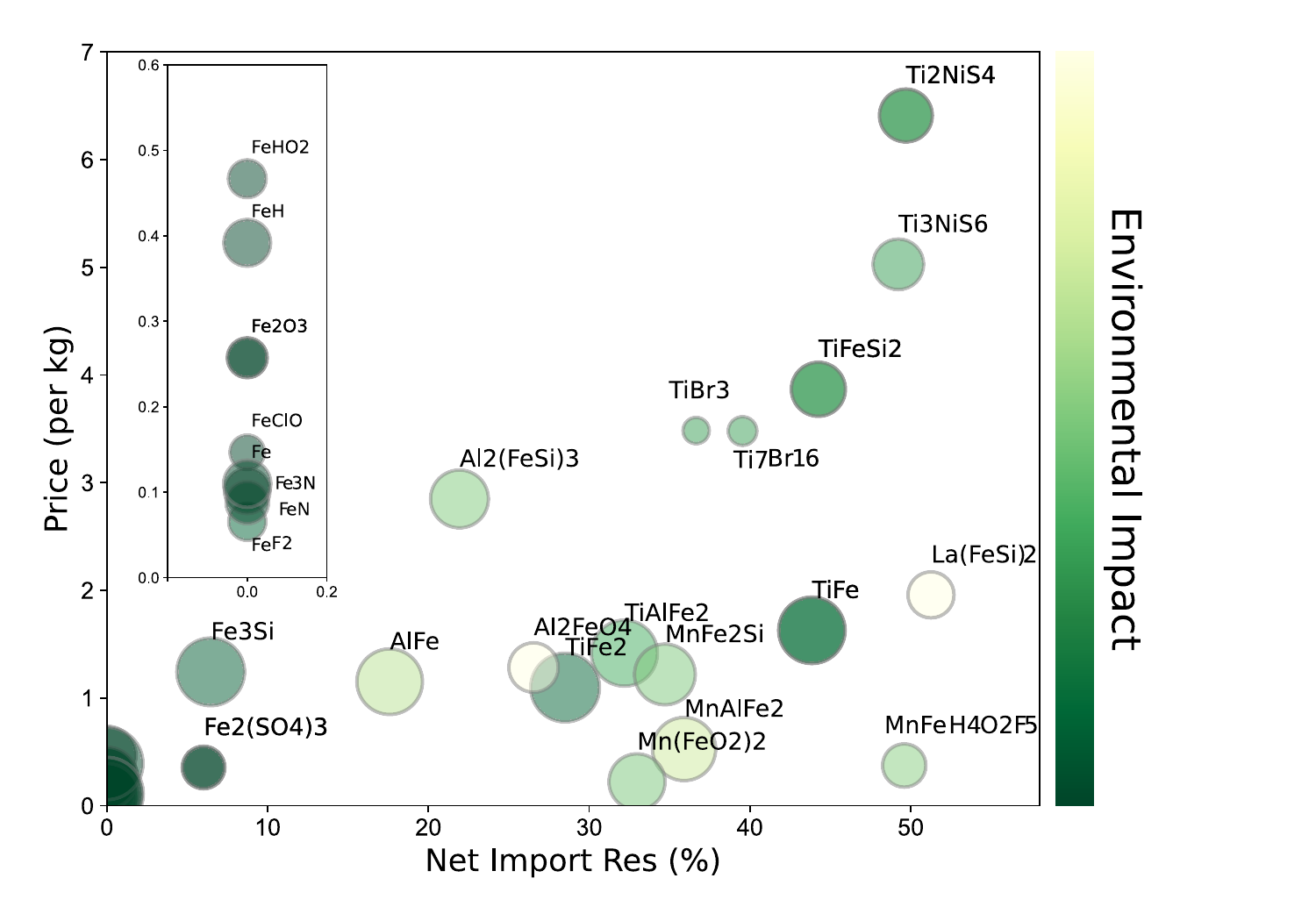}
 \caption{\textbf{\textbf{Topological materials ranking based on the price of the raw materials and the net import resilience (NIR).}} Price is in USD per kg, and color indicates the Env score and the size of the circle scales with the recycle rate of the raw materials.}
 \label{figure4}
\end{figure}

%\section*{Results}\label{sec2}

The screening process reduces the pool of topological materials from over 16,000 candidates to roughly 200. As shown in Figure \ref{figure2}, there is significant overlap between materials that are low-cost, and those with a low environmental impact score. In particular, Earth-abundant elements such as Fe, Ni, and Si, as well as other commonly used transition metals and group III-V elements are both cheap and have low environmental impact. 

These suitable elements are also commonly shared among topological materials. Figure \ref{figure3} shows more details on these elements and the impact of the filtering process on the total number of topological materials candidates. In Figure \ref{figure3}a, a comprehensive depiction of element-specific pricing, NIR, and Env scores is presented. Note that the Env score of an element is derived from the amalgamation of EGov and cumulative energy demand \cite{EnvCri}, which is further scaled by the recycling rate. As a result, elements entailing higher environmental impact during the production phase, such as Co, might exhibit higher Env scores compared to elements with relatively lower environmental impact. As another interesting case, Ti is very low cost, but has a high NIR value, implying that it is susceptible to supply chain disruptions. Moreover, Figure \ref{figure3}b elucidates the count of topological insulators and semimetals contained within the initial topological materials database prior to and post the rigorous screening process. When sustainability is taken into account, topological materials that include Mn, Co, and Al are not as prominent in the database. Meanwhile, topological materials containing Fe and Ti become more prominent in the sustainability database. 

For the purpose of ranking, a comprehensive scoring system is employed, involving the calculation of weighted averages from the individual scores assigned to the pure constituent materials within each compound. To visually represent this ranking, Figure \ref{figure4} showcases the top 36 materials arranged according to their standings based on price and NIR. Based on this plot, materials which are environmentally friendly also tend to have lower prices and NIR scores. Such correlation is expected based on the the fact that difficult-to-extract raw materials require more energy resources and often occur outside of the domestic supply-chain, yielding higher prices. More abundant elements, like Fe, are cheaper and easier to refine. 

Table \ref{tab1} lists the top-ranked 244 of the topological materials database when factoring in scores for price, melting temperature of the constituent elements, and Env (the full database can be found in the Supplementary Information as Table \ref{tab3}). The price of the raw materials is less than 50\$/kg without any toxic elements as these materials mainly contain common elements such as Si, Ni, and Fe. While the high-throughput screening through ab initio calculations and sustainability scores reveals a handful of materials which are unlikely to be topological (such as elemental metals), most possess prior reports of synthesis and notably, some have even been reported in recent years to exhibit experimental signatures that are exemplary of topological physics. We highlight some examples from Table \ref{tab1}.

\textit{Co$_3$Sn$_2$S$_2$}. Co$_3$Sn$_2$S$_2$ has been demonstrated to be a magnetic Weyl semimetal where upon the magnetic properties arises from the kagom\'e-lattice of the Co layer at a transition temperature of 175 K \cite{wang2018,morali2019}. The materials are synthesized in single-crystalline form via flux method using the respective elemental powders and using a Sn flux, but can also be grown in thin film form using magnetron sputtering \cite{fujiwara2019} or molecular beam epitaxy \cite{li2020thin}. From a topological physics point of view, the compound exhibits a large intrinsic anomalous Hall effect (AHE) that originates from the presence of topological states. As a result, Co$_3$Sn$_2$S$_2$ may serve as a material platform in novel spintronic devices that are based on the AHE \cite{wang2023}.

\textit{CoSi}. CoSi is a chiral topological material which has gained interest due to its resistivity scaling in nanostructures and compatibility with CMOS technology. It is a member of the B20 family that crystallizes in a simple cubic structure which has been well-studied for its topological multifold band degeneracies and long Fermi arcs that extend from the zone center to the corner of the Brillouin zone \cite{rao2019}. The material is usually synthesized in a chemical vapor transport method with the constituent Co and Si powders and with iodine gas as the transport agent. There is interest from the semiconductor industry in the exploration of using CoSi as post-copper interconnects in back-end-of-line processing due to its unconventional resistivity scaling \cite{lien2023}.

\textit{Co$_2$MnX (X = Al, Si)}. These compounds have attracted attention for their unique thermoelectric transport properties. They are formed in the L2$_1$ phase of the cubic space group via the floating-zone technique and host topologically nontrivial states. In these materials, the presence of topological nodal points near the Fermi level is expected to give rise to exotic transverse transport properties such as anomalous Hall and Nernst effects due to the presence of large Berry curvature. Indeed, Co$_2$MnAl has been demonstrated to be a ferromagnetic topological semimetal that exhibits a large room temperature anomalous Hall effect and a corresponding large Hall angle \cite{li2020} whereas thin films of Co$_2$MnAl showcase an enhanced anomalous Nernst effect upon doping with Si to shift the position of the Fermi level \cite{sakuraba2020}. As a result of the anomalous Hall and Nernst effects, these materials may serve as platforms for next-generation spintronic and thermoelectric devices, respectively.

\textit{YMn$_6$Sn$_6$}. Similar to Co$_3$Sn$_2$S$_2$, YMn$_6$Sn$_6$ features ferromagnetic kagom\'e layers which also result in an intrinsic anomalous Hall effect from Berry curvature and strong electron correlations originating from the presence of yttrium and flat bands \cite{ghimire2020,li2021}. The compound is synthesized in single-crystalline form through a flux method involving Sn flux. Kagom\'e-net magnets such as YMn$_6$Sn$_6$ offer a material system in which one can explore how topological physics can manipulate the magnetic properties resulting exotic magnetic phases. Notably, in the case of YMn$_6$Sn$_6$, observations of topological Hall effect showcase exotic magnetic phases involving magnetic skyrmions which may serve as candidate bits for future data-storage and spintronic devices.

Some of the most well-known and actively-studied topological materials have low sustainability scores. Table \ref{tab2} shows the scoring for a subset of these materials. With the exception of Co$_3$Sn$_2$S$_2$ which appears in both Table \ref{tab1} and \ref{tab2}, all of these archetypal topological materials possess an Env above 2 or contain toxic elements in the form of precursors of phosphorus and arsenic. Some of the materials in Table \ref{tab2} also contain elements with high individual melting temperatures or vapor pressures. These factors would make synthesis in an industrial setting difficult, but not impossible. In contrast to these established topological materials, a disposition toward research on topological materials with superior environmental and industrial metrics is feasible, as shown by the examples above.

\section*{Discussion}

There are several noteworthy factors that our filtering and scoring procedure does not take into account. A significant portion of the materials listed in the topological material database have never been synthesized. As a result, the precise raw materials required for synthesizing these materials cannot be obtained solely through data mining for the synthesizing recipe. In the context of our analysis, it is assumed that the raw materials requisite for each topological material are composed of pure elements as stipulated within their respective chemical formulas. However, it may be the case that other precursor compounds are required for particular kinds of synthesis. Moreover, some materials may be synthesizable despite having $E_{hull}>0$ eV within the MaterialsProject database. One example of this case is Mn$_3$Sn, which the database claims has $E_{hull}=0.18$ eV but is a frequently studied topological material. 

Furthermore, it is an outstanding problem for high throughput studies to filter for anticipated technical performance. Taking topological materials' performance into account requires a quantifiable metric. For topological semimetals, such an index could be the distance of topological features (e.g. band-crossings) from the Fermi level, and for topological insulators it could be the size of the topological band-gap. These metrics could be extracted from high throughput calculations, but such a method is beyond the scope of the current work. Alternatively, performance metrics could be application specific. For example, in thermoelectric performance the magnitude of the anomalous Nernst affect can be taken into account. However, many of these values are experimentally verified, and require a large amount of effort to attain. A lack of topological performance metrics helps explain why our database includes elemental materials like Fe, Ni, Ti, and Sn. These materials satisfy the constraints imposed by the Topological Quantum Chemistry Database, but it is not immediately clear that their macroscopic properties are influenced by this classification. 
 
It may also be valuable to consider more geographically localized sustainability analysis. Here, a primary basis for the sustainability score comes from global raw-materials extraction. More detailed analysis could focus on specific elements that are extracted in specific locales. The concentration of production in a few countries for key raw materials raises concerns about supply chain vulnerabilities. Notable examples are the Democratic Republic of the Congo and China, where the mining and refinement of cobalt is concentrated. Cobalt is an important raw material for the production of rechargeable batteries, and our work also shows that it frequently appears in sustainable topological materials. In 2022, 76$\%$ of the American cobalt consumption relied on imports from these two countries \cite{Fortier2018}. As a result, disruptions in countries such as these can lead to shortages and price fluctuations, affecting the entire supply chain. Moreover, the NIR score is based on data specific to the United States, and will need to be modified for analyses specific to other locales. 

There is a strong incentive for the topological materials research field --- which currently focuses on several heavily studied materials --- to invest more time and effort toward studying new sustainable topological materials. These materials also tend to be cheaper and more resilient to supply-chain disruptions, giving them potential for adoption within a broader industrial framework. Enveloping these considerations into the materials discovery and design process more generally will not only improve the potential technological impact of those discoveries but also assist in establishing an environmentally sustainable, resilient, and green industrial economy.

\onecolumn

\captionof{table}{List of parameters for the top 200 highly ranked topological materials by sustainability. E$_F$: Fermi level in electronvolts (eV). Price is in USD (in 2023) per kilogram (\$/kg). E$_\text{hull}$: energy above the complex hull; NIR: net import resilience; K$_{\text{voigt}}$: Bulk modulus (Voigt average, GPa); E$_{\text{g, D}}$: direct band gap in electronvolts; E$_{\text{g, ID}}$: indirect band gap in electronvolts; SG: space group number index; Env: environmental score. The materials are categorized (C) into three groups, with spin-orbit coupling taken into account: (1) topological insulator, (2) semimetal, and (3) trivial.}
\label{tab1}
\resizebox{\textwidth}{!}{\begin{tabular}{c||c||c|c|c|c|c|c|c|c|c|c|c}\toprule\multicolumn{1}{l||}{\textbf{ID}} & \multicolumn{1}{l||}{\textbf{Formula}} & \multicolumn{1}{l|}{\textbf{E$_{F}$}} & \multicolumn{1}{l|}{\textbf{Price}} & \multicolumn{1}{l|}{\textbf{E$_\text{hull}$}} & \multicolumn{1}{l|}{\textbf{NIR}} & \multicolumn{1}{l|}{\textbf{K$_\text{voigt}$}} & \multicolumn{1}{l|}{\textbf{E$_{g, \text{D}}$}} & \multicolumn{1}{l|}{\textbf{E$_{g, \text{ID}}$}} & \multicolumn{1}{l|}{\textbf{Symmetry}} & \multicolumn{1}{l|}{\textbf{SG}} & \multicolumn{1}{l|}{\textbf{Env}} & \multicolumn{1}{l}{\textbf{C}} \\ \midrule
mp-13&Fe&5.30&0.11&0.00&0.00&182.46&0.000&0.000&Cubic&229&1.15&2\\ 
mp-23&Ni&8.06&25.00&0.00&56.00&198.07&0.000&0.000&Cubic&225&1.32&2\\ 
mp-72&Ti&5.89&3.40&0.00&95.00&113.20&0.043&0.000&Hexagonal&191&1.35&1\\ 
mp-117&Sn&4.91&35.20&0.00&77.00&38.49&0.000&0.000&Cubic&227&1.59&2\\ 
mp-246&TiF$_3$&1.80&1.55&0.00&95.00&103.68&0.001&0.000&Cubic&221&1.33&2\\ 
mp-284&AlCo&8.42&44.87&0.00&69.09&178.60&0.059&0.000&Cubic&221&1.93&1\\ 
mp-305&TiFe&6.69&1.63&0.00&43.85&193.94&0.009&0.000&Cubic&221&1.24&1\\ 
mp-351&SiNi&10.03&19.50&0.00&52.44&165.59&0.044&0.000&Orthorhombic&62&1.38&1\\ 
mp-458&Ti$_2$O$_3$&7.03&2.46&0.00&95.00&215.86&0.007&0.000&Trigonal&167&1.35&2\\ 
mp-492&TiN&8.77&2.63&0.00&95.00&280.94&0.000&0.000&Cubic&225&1.35&2\\ 
mp-633&LaSn$_3$&8.90&25.60&0.00&82.05&-&0.000&0.000&Cubic&221&1.79&2\\ 
mp-823&TiCo&6.31&36.78&0.00&84.52&163.17&0.000&0.000&Cubic&221&1.56&2\\ 
mp-828&SiNi$_3$&9.51&22.66&0.00&54.49&-&0.069&0.000&Cubic&221&1.34&1\\ 
mp-943&Co$_3$S$_4$&6.14&37.09&0.00&49.51&-&0.000&0.000&Cubic&227&1.73&2\\ 
mp-1048&TiNi&7.01&15.30&0.00&73.52&-&0.013&0.000&Monoclinic&11&1.33&1\\ 
mp-1057&Al$_3$Ni$_2$&7.84&16.15&0.00&55.18&130.42&0.000&0.000&Trigonal&164&1.74&2\\ 
mp-1118&SiNi$_2$&10.12&21.72&0.00&53.88&197.57&0.038&0.000&Orthorhombic&62&1.35&1\\ 
mp-1215&Ti$_2$O&7.52&3.00&0.00&95.00&184.03&0.001&0.000&Trigonal&164&1.35&2\\ 
mp-1364&YNi&4.79&35.84&0.00&82.50&78.76&0.001&0.000&Orthorhombic&62&1.32&2\\ 
mp-1409&TiNi$_3$&8.17&20.38&0.00&64.34&192.48&0.010&0.000&Hexagonal&194&1.33&2\\ 
mp-1418&FeNi$_3$&7.50&19.01&0.00&42.52&203.11&0.000&0.000&Cubic&221&1.28&2\\ 
mp-1431&MnSi&9.24&2.71&0.00&81.39&209.61&0.000&0.000&Cubic&198&1.87&2\\ 
mp-1487&AlNi&10.48&18.17&0.00&55.37&161.58&0.000&0.000&Cubic&221&1.64&2\\ 
mp-1513&Co$_9$S$_8$&6.30&43.12&0.00&55.46&-&0.000&0.000&Cubic&225&1.73&2\\ 
mp-1570&YFe$_2$&4.55&19.12&0.00&44.32&-&0.000&0.000&Cubic&227&1.15&2\\ 
mp-1804&Fe$_3$N&6.24&0.10&0.00&0.00&206.76&0.002&0.000&Hexagonal&182&1.15&2\\ 
mp-1808&Ti$_2$Ni&6.57&11.61&0.00&80.18&141.42&0.000&0.000&Cubic&227&1.34&2\\ 
mp-1823&Ti$_3$Al&6.27&3.39&0.00&88.51&115.31&0.000&0.000&Hexagonal&194&1.51&2\\ 
mp-1824&NdNi$_5$&6.47&35.71&0.00&68.85&-&0.000&0.000&Hexagonal&191&1.65&2\\ 
mp-1953&TiAl&6.81&3.37&0.00&80.22&115.06&0.000&0.000&Tetragonal&123&1.71&2\\ 
mp-1977&NdSn$_3$&7.76&41.63&0.00&82.19&-&0.003&0.000&Cubic&221&1.80&1\\ 
mp-2033&Ni$_3$N&8.16&23.16&0.00&56.00&197.04&0.004&0.000&Hexagonal&182&1.32&2\\ 
mp-2056&Ni$_3$Se$_2$&6.13&23.58&0.00&53.16&105.87&0.017&0.000&Trigonal&155&1.79&2\\ 
mp-2070&CoS$_2$&8.01&30.68&0.00&43.17&124.28&0.000&0.000&Cubic&205&1.73&2\\ 
mp-2090&FeCo&5.86&32.86&0.00&39.02&189.18&0.000&0.000&Cubic&221&1.45&2\\ 
mp-2108&Ti$_5$Si$_3$&7.39&4.60&0.00&81.98&139.78&0.013&0.000&Hexagonal&193&1.39&1\\ 
mp-2152&YNi$_5$&6.50&29.19&0.00&66.23&148.21&0.000&0.000&Hexagonal&191&1.32&2\\ 
mp-2199&Fe$_3$Si&7.73&1.24&0.00&6.46&211.88&0.000&0.000&Cubic&225&1.20&2\\ 
mp-2213&FeNi&6.84&12.87&0.00&28.70&186.74&0.001&0.000&Tetragonal&123&1.24&2\\ 
mp-2291&Si$_2$Ni&10.02&16.69&0.00&50.62&150.26&0.002&0.000&Cubic&225&1.41&2\\ 
mp-2317&LaNi$_5$&8.43&17.29&0.00&68.53&135.74&0.000&0.000&Hexagonal&191&1.64&2\\ 
mp-2379&CoSi$_2$&9.49&36.62&0.00&60.87&176.91&0.000&0.000&Cubic&225&1.62&2\\ 
mp-2454&TiFe$_2$&5.79&1.10&0.00&28.50&146.75&0.003&0.000&Hexagonal&194&1.21&2\\ 
mp-2538&Y$_5$Si$_3$&4.46&37.42&0.00&91.24&-&0.000&0.000&Hexagonal&193&1.50&2\\ 
mp-2591&Ti$_3$O&6.73&3.12&0.00&95.00&-&0.000&0.000&Trigonal&163&1.35&2\\ 
mp-2593&AlNi$_3$&8.51&22.12&0.00&55.73&179.51&0.000&0.000&Cubic&221&1.46&2\\ 
mp-2658&AlFe&8.18&1.15&0.00&17.59&174.59&0.000&0.000&Cubic&221&1.54&2\\ 
mp-3337&Ti$_6$Si$_7$Ni$_1$$_6$&8.79&18.29&0.00&62.35&-&0.008&0.000&Cubic&225&1.35&2\\ 
mp-3489&Nd(FeSi)$_2$&6.23&28.05&0.00&52.01&-&0.023&0.000&Tetragonal&139&1.75&2\\ 
mp-3602&YAl$_4$Ni&6.95&22.10&0.00&70.46&92.59&0.000&0.000&Orthorhombic&63&1.99&2\\ 
mp-3623&MnAlCo$_2$&6.61&38.15&0.00&79.63&191.80&0.006&0.000&Cubic&225&1.90&1\\ 
mp-3657&TiCo$_2$Si&8.10&40.86&0.00&76.20&201.33&0.000&0.000&Cubic&225&1.60&2\\ 
\bottomrule\end{tabular}}

\clearpage
\resizebox{\textwidth}{!}{\begin{tabular}{c||c||c|c|c|c|c|c|c|c|c|c|c}\toprule\multicolumn{1}{l||}{\textbf{ID}} & \multicolumn{1}{l||}{\textbf{Formula}} & \multicolumn{1}{l|}{\textbf{E$_{F}$}} & \multicolumn{1}{l|}{\textbf{Price}} & \multicolumn{1}{l|}{\textbf{E$_\text{hull}$}} & \multicolumn{1}{l|}{\textbf{NIR}} & \multicolumn{1}{l|}{\textbf{K$_\text{voigt}$}} & \multicolumn{1}{l|}{\textbf{E$_{g, \text{D}}$}} & \multicolumn{1}{l|}{\textbf{E$_{g, \text{ID}}$}} & \multicolumn{1}{l|}{\textbf{Symmetry}} & \multicolumn{1}{l|}{\textbf{SG}} & \multicolumn{1}{l|}{\textbf{Env}} & \multicolumn{1}{l}{\textbf{C}} \\ \midrule
mp-3854&Y(MnSi)$_2$&6.43&16.76&0.00&87.88&127.62&0.000&0.000&Tetragonal&139&1.87&2\\ 
mp-4007&Nd(SiNi)$_2$&6.95&36.75&0.00&71.76&-&0.022&0.000&Tetragonal&139&1.80&1\\ 
mp-4088&La(FeSi)$_2$&8.10&1.96&0.00&51.26&-&0.000&0.000&Tetragonal&139&1.74&2\\ 
mp-4196&NdFeSi$_2$&6.38&34.14&0.00&63.34&-&0.004&0.000&Orthorhombic&63&1.88&1\\ 
mp-4492&MnCo$_2$Si&7.54&38.61&0.00&78.23&219.29&0.000&0.000&Cubic&225&1.79&2\\ 
mp-4577&Y$_3$(AlNi$_3$)$_2$&6.06&30.40&0.00&73.28&-&0.000&0.000&Cubic&229&1.46&2\\ 
mp-4656&Y(Al$_2$Fe)$_4$&7.17&8.64&0.00&38.91&123.10&0.000&0.000&Tetragonal&139&1.74&2\\ 
mp-4847&Y$_4$Al$_2$$_3$Ni$_6$&7.47&19.68&0.00&66.85&-&0.000&0.000&Monoclinic&12&1.97&2\\ 
mp-4922&MnAlNi$_2$&7.58&15.17&0.00&67.86&-&0.000&0.000&Cubic&225&1.66&2\\ 
mp-4978&Ti$_2$AlN&7.41&3.03&0.00&85.99&155.37&0.010&0.000&Hexagonal&194&1.57&1\\ 
mp-5063&YAl$_3$Ni$_2$&6.96&24.46&0.00&69.06&113.28&0.001&0.000&Hexagonal&191&1.74&2\\ 
mp-5129&Y(CoSi)$_2$&6.71&44.89&0.00&77.49&145.19&0.000&0.000&Tetragonal&139&1.66&2\\ 
mp-5176&Y(SiNi)$_2$&7.12&27.46&0.00&68.55&136.24&0.000&0.000&Tetragonal&139&1.38&2\\ 
mp-5288&Y(FeSi)$_2$&6.45&16.69&0.00&44.47&139.90&0.000&0.000&Tetragonal&139&1.27&2\\ 
mp-5407&TiAlCo$_2$&7.10&40.39&0.00&77.64&177.70&0.000&0.000&Cubic&225&1.72&2\\ 
mp-5423&Pr(MnSi)$_2$&6.26&48.75&0.00&87.64&-&0.000&0.000&Tetragonal&139&1.97&2\\ 
mp-5526&La(CoSi)$_2$&8.23&25.95&0.00&78.87&-&0.000&0.000&Tetragonal&139&1.95&2\\ 
mp-5528&Ni$_3$(SnS)$_2$&8.03&26.73&0.00&60.66&-&0.016&0.000&Trigonal&166&1.47&1\\ 
mp-5529&MnFe$_2$Si&8.05&1.22&0.00&34.71&232.87&0.000&0.000&Cubic&225&1.46&2\\ 
mp-5627&Pr(FeSi)$_2$&6.29&48.50&0.00&51.54&-&0.000&0.000&Tetragonal&139&1.65&2\\ 
mp-5898&La(SiNi)$_2$&8.69&11.27&0.00&71.36&-&0.000&0.000&Tetragonal&139&1.79&2\\ 
mp-6988&FeN&5.56&0.09&0.00&0.00&267.00&0.000&0.000&Cubic&216&1.15&2\\ 
mp-7030&LaSiNi&8.04&8.11&0.00&78.64&-&0.000&0.000&Tetragonal&109&1.95&2\\ 
mp-7092&TiSi&8.19&5.10&0.00&76.51&146.43&0.004&0.000&Orthorhombic&62&1.41&1\\ 
mp-7186&NdAl$_3$Ni$_2$&6.53&33.56&0.00&71.95&-&0.000&0.000&Hexagonal&191&1.98&2\\ 
mp-7187&TiAlNi$_2$&8.02&16.58&0.00&65.43&161.85&0.000&0.000&Cubic&225&1.47&2\\ 
mp-7577&CoSi&9.64&45.86&0.00&65.99&210.51&0.000&0.000&Cubic&198&1.66&2\\ 
mp-8282&Ti$_2$N&7.00&2.97&0.00&95.00&204.11&0.000&0.000&Tetragonal&136&1.35&2\\ 
mp-9177&Y$_2$Al$_3$Si$_2$&6.62&26.56&0.00&78.37&86.29&0.000&0.000&Monoclinic&12&2.00&2\\ 
mp-9972&YSi&5.55&34.60&0.00&86.80&89.42&0.000&0.000&Orthorhombic&63&1.50&2\\ 
mp-10010&Al(CoSi)$_2$&9.15&40.15&0.00&64.38&177.25&0.000&0.000&Trigonal&164&1.75&2\\ 
mp-10527&Y$_2$AlSi$_2$&5.72&31.36&0.00&83.41&87.57&0.000&0.000&Orthorhombic&71&1.77&2\\ 
mp-10884&AlFeCo$_2$&6.61&38.00&0.00&51.90&187.93&0.000&0.000&Cubic&225&1.65&2\\ 
mp-11352&PrAl$_5$Ni$_2$&7.20&45.51&0.00&69.29&-&0.015&0.000&Orthorhombic&71&1.95&1\\ 
mp-11356&La$_3$Co$_2$Sn$_7$&8.49&27.24&0.00&82.41&-&0.000&0.000&Orthorhombic&65&1.82&2\\ 
mp-11385&YFe$_5$&4.49&10.47&0.00&24.15&107.81&0.000&0.000&Hexagonal&191&1.15&2\\ 
mp-11501&MnNi$_3$&7.24&19.06&0.00&66.46&174.08&0.000&0.000&Cubic&221&1.49&2\\ 
mp-11726&LaSi$_4$Ni$_9$&9.38&18.27&0.00&61.36&-&0.000&0.000&Tetragonal&140&1.52&2\\ 
mp-11765&Ti$_5$Se$_4$&6.14&13.98&0.00&69.40&-&0.004&0.000&Tetragonal&87&1.90&1\\ 
mp-12613&LaNiSn&8.80&18.29&0.00&81.01&-&0.000&0.000&Orthorhombic&62&1.85&2\\ 
mp-12701&NdSn$_2$&6.55&43.63&0.00&83.80&-&0.006&0.000&Orthorhombic&65&1.86&1\\ 
mp-13010&YSn$_2$&6.86&37.33&0.00&83.27&59.13&0.000&0.000&Orthorhombic&63&1.59&2\\ 
mp-13093&La$_3$Ni$_2$Sn$_7$&8.77&23.88&0.00&80.69&-&0.000&0.000&Orthorhombic&65&1.78&2\\ 
mp-13094&YAl$_2$Ni&6.24&27.13&0.00&74.87&92.60&0.000&0.000&Orthorhombic&63&1.81&2\\ 
mp-13095&YAlNi&5.77&30.81&0.00&78.10&93.21&0.000&0.000&Hexagonal&189&1.64&1\\ 
mp-13993&Ti$_3$NiS$_6$&6.23&5.03&0.00&49.23&71.95&0.004&0.000&Trigonal&148&1.34&1\\ 
mp-16514&Al$_3$Ni$_5$&8.91&20.31&0.00&55.57&167.25&0.000&0.000&Orthorhombic&65&1.54&2\\ 
mp-16515&Al$_4$Ni$_3$&8.41&16.76&0.00&55.24&-&0.001&0.000&Cubic&230&1.71&1\\ 
mp-16744&Y(Al$_5$Fe)$_2$&7.51&10.05&0.00&49.87&-&0.000&0.000&Orthorhombic&63&2.00&2\\ 
mp-17114&Nd$_2$Sn$_2$O$_7$&3.87&39.21&0.00&86.87&-&0.002&0.002&Cubic&227&1.98&1\\ 
mp-18554&Mn$_4$Be$_3$Si$_3$SeO$_1$$_2$&2.96&36.98&0.00&72.47&-&0.000&0.000&Cubic&218&2.00&2\\ 
mp-18583&Ti$_4$Si$_7$Ni$_4$&9.30&12.99&0.00&64.52&-&0.008&0.000&Tetragonal&139&1.39&1\\ 
mp-18706&Y$_5$(FeTe)$_2$&5.07&45.59&0.00&78.37&-&0.000&0.000&Orthorhombic&63&1.96&2\\ 
mp-18732&TiNiO$_3$&4.92&10.73&0.00&73.52&179.63&0.058&0.000&Trigonal&148&1.33&1\\ 
mp-18750&Mn(FeO$_2$)$_2$&4.19&0.22&0.00&32.97&157.98&0.000&0.000&Cubic&227&1.45&2\\ 
mp-19009&NiO&5.87&19.77&0.00&56.00&181.29&0.000&0.000&Cubic&225&1.32&2\\ 
mp-19072&Si(NiO$_2$)$_2$&5.53&15.27&0.00&53.88&-&0.004&0.000&Orthorhombic&62&1.35&1\\ 
mp-19082&TiMnO$_3$&3.64&1.27&0.00&97.67&162.08&0.015&0.000&Trigonal&148&1.73&1\\ 
mp-19314&Mn$_4$Be$_3$Si$_3$SO$_1$$_2$&1.95&36.99&0.00&72.11&-&0.002&0.000&Cubic&218&1.92&2\\ 
mp-19379&CoSO$_4$&1.77&24.57&0.00&53.80&-&0.000&0.000&Orthorhombic&62&1.73&2\\ 
mp-19442&Mn(Ni$_3$O$_4$)$_2$&4.80&16.60&0.00&61.94&159.91&0.000&0.000&Cubic&225&1.42&2\\ 
mp-19448&Ni$_3$TeO$_6$&6.18&33.51&0.00&63.98&157.51&0.004&0.000&Trigonal&146&1.74&2\\ 
mp-19770&Fe$_2$O$_3$&2.88&0.26&0.00&0.00&193.04&0.012&0.000&Trigonal&167&1.15&1\\ 
\bottomrule\end{tabular}}

\clearpage
\resizebox{\textwidth}{!}{\begin{tabular}{c||c||c|c|c|c|c|c|c|c|c|c|c}\toprule\multicolumn{1}{l||}{\textbf{ID}} & \multicolumn{1}{l||}{\textbf{Formula}} & \multicolumn{1}{l|}{\textbf{E$_{F}$}} & \multicolumn{1}{l|}{\textbf{Price}} & \multicolumn{1}{l|}{\textbf{E$_\text{hull}$}} & \multicolumn{1}{l|}{\textbf{NIR}} & \multicolumn{1}{l|}{\textbf{K$_\text{voigt}$}} & \multicolumn{1}{l|}{\textbf{E$_{g, \text{D}}$}} & \multicolumn{1}{l|}{\textbf{E$_{g, \text{ID}}$}} & \multicolumn{1}{l|}{\textbf{Symmetry}} & \multicolumn{1}{l|}{\textbf{SG}} & \multicolumn{1}{l|}{\textbf{Env}} & \multicolumn{1}{l}{\textbf{C}} \\ \midrule
mp-19807&Co$_3$(SnS)$_2$&8.13&41.11&0.00&68.05&97.08&0.000&0.000&Trigonal&166&1.65&2\\ 
mp-19963&TiFe$_2$Sn&8.83&15.65&0.00&49.19&181.40&0.000&0.000&Cubic&225&1.37&2\\ 
mp-20000&Nd(NiSn)$_2$&7.18&39.25&0.00&77.26&-&0.004&0.000&Tetragonal&129&1.73&2\\ 
mp-20112&Ni$_3$Sn&9.53&29.11&0.00&64.46&164.45&0.002&0.000&Hexagonal&194&1.43&2\\ 
mp-20174&Ni$_3$Sn$_4$&9.22&32.44&0.00&71.32&99.05&0.053&0.000&Monoclinic&12&1.51&1\\ 
mp-20211&Mn$_3$Si&7.94&1.17&0.00&91.99&206.25&0.000&0.000&Cubic&225&1.97&2\\ 
mp-20311&FeSe&4.01&12.93&0.00&29.29&29.57&0.018&0.000&Tetragonal&129&1.83&1\\
mp-20382&Ti$_6$Sn$_5$&8.23&24.83&0.00&82.87&104.98&0.001&0.000&Hexagonal&194&1.51&2\\ 
mp-20536&CoSn&8.02&44.72&0.00&76.67&126.54&0.000&0.000&Hexagonal&191&1.64&2\\ 
mp-20557&YSiNi&6.28&31.39&0.00&76.51&113.01&0.001&0.000&Orthorhombic&62&1.38&2\\ 
mp-20690&NdFeSi&5.29&37.36&0.00&65.60&-&0.006&0.000&Tetragonal&129&1.93&1\\ 
mp-20840&MnCo$_2$Sn&7.82&40.17&0.00&80.93&169.11&0.000&0.000&Cubic&225&1.73&2\\ 
mp-20847&Ti$_5$Sn$_3$&7.80&22.42&0.00&84.23&107.12&0.006&0.000&Hexagonal&193&1.49&1\\ 
mp-20949&MnSiNi&8.51&11.94&0.00&70.88&166.66&0.000&0.000&Orthorhombic&62&1.64&2\\ 
mp-21249&YMnSi&5.44&23.54&0.00&91.02&91.66&0.014&0.000&Tetragonal&129&1.87&1\\ 
mp-21260&FeSn&7.71&23.97&0.00&52.37&100.27&0.014&0.000&Hexagonal&191&1.45&2\\ 
mp-21306&TiCoSi&7.58&30.79&0.00&76.29&181.36&0.000&0.000&Orthorhombic&62&1.55&2\\ 
mp-21437&Fe$_2$TeO$_6$&3.39&26.85&0.00&39.99&-&0.012&0.000&Tetragonal&136&1.77&1\\ 
mp-21467&TiCo$_2$Sn&8.27&41.74&0.00&79.61&168.18&0.000&0.000&Cubic&225&1.61&1\\ 
mp-21606&TiMnSi$_2$&8.75&3.85&0.00&79.06&-&0.002&0.000&Orthorhombic&55&1.65&1\\ 
mp-21661&YSi$_3$Ni$_5$&8.19&25.36&0.00&62.40&-&0.000&0.000&Orthorhombic&62&1.36&2\\ 
mp-21662&TiFeSi$_2$&8.81&3.87&0.00&44.25&180.93&0.006&0.000&Orthorhombic&55&1.33&1\\ 
mp-21856&Co$_2$SiO$_4$&4.63&37.13&0.00&70.03&-&0.002&0.000&Orthorhombic&62&1.69&1\\ 
mp-22179&YTiSi&5.17&25.54&0.00&89.18&89.51&0.000&0.000&Tetragonal&129&1.41&2\\ 
mp-22260&Y(MnSn)$_6$&8.23&25.55&0.00&85.51&110.13&0.000&0.000&Hexagonal&191&1.73&2\\ 
mp-22290&Mn$_6$Si$_7$Ni$_1$$_6$&8.30&17.10&0.00&64.42&-&0.003&0.000&Cubic&225&1.51&2\\ 
mp-22617&YNiSn&6.69&35.56&0.00&80.05&91.00&0.002&0.000&Orthorhombic&62&1.50&2\\ 
mp-22658&Co$_2$NiS$_4$&6.30&29.58&0.00&45.64&-&0.032&0.000&Cubic&227&1.60&1\\ 
mp-23069&Fe(IO$_3$)$_3$&2.97&27.05&0.00&43.60&-&0.000&0.000&Hexagonal&173&1.15&2\\ 
mp-23853&CoH$_4$(ClO)$_2$&-0.28&23.21&0.00&76.00&-&0.000&0.000&Monoclinic&12&1.73&2\\ 
mp-24719&NiH&7.87&24.85&0.00&56.00&199.62&0.000&0.000&Cubic&225&1.32&2\\ 
mp-24726&TiH$_2$&5.74&3.91&0.00&95.00&142.05&0.016&0.000&Tetragonal&139&1.35&1\\ 
mp-27276&Si$_1$$_2$Ni$_3$$_1$&9.46&22.34&0.00&54.28&198.91&0.001&0.000&Trigonal&150&1.35&2\\ 
mp-27848&TiIN&3.70&28.42&0.00&62.32&43.70&0.026&0.000&Orthorhombic&59&1.35&1\\ 
mp-28137&Co(ClO$_4$)$_2$&-0.84&14.90&0.00&76.00&-&0.000&0.000&Trigonal&148&1.73&2\\ 
mp-28214&TiBr$_3$&1.74&3.48&0.00&36.65&-&0.014&0.000&Triclinic&2&1.35&1\\ 
mp-28231&Ti$_2$FeO$_5$&2.93&1.64&0.00&60.00&167.94&0.006&0.000&Orthorhombic&63&1.28&1\\ 
mp-29110&Al$_2$(FeSi)$_3$&9.19&2.85&0.00&21.93&-&0.006&0.000&Triclinic&2&1.46&1\\ 
mp-30033&CoBr$_2$&0.88&19.77&0.00&38.74&17.17&0.000&0.000&Trigonal&164&1.73&2\\ 
mp-30084&Al$_2$FeO$_4$&7.21&1.28&0.00&26.54&184.70&0.008&0.000&Cubic&227&1.74&2\\ 
mp-30875&Ti$_2$Sn&8.05&21.00&0.00&85.04&115.41&0.010&0.000&Hexagonal&194&1.48&1\\ 
mp-31080&YAlSi&6.32&28.73&0.00&80.65&88.50&0.050&0.000&Orthorhombic&63&1.91&1\\ 
mp-31185&MnAlFe$_2$&6.99&0.53&0.00&35.90&196.03&0.000&0.000&Cubic&225&1.57&2\\ 
mp-31187&TiAlFe$_2$&6.95&1.42&0.00&32.19&184.56&0.000&0.000&Cubic&225&1.37&2\\ 
mp-504722&Ni(NO$_3$)$_2$&0.09&8.35&0.00&56.00&-&0.000&0.000&Cubic&205&1.32&2\\ 
mp-504945&MnFeH$_4$O$_2$F$_5$&-1.02&0.37&0.00&49.59&-&0.007&0.000&Orthorhombic&74&1.47&1\\ 
mp-505332&LaFeSi&7.82&1.66&0.00&64.89&-&0.000&0.000&Tetragonal&129&1.92&2\\ 
mp-505405&TiMnH$_1$$_2$(OF)$_6$&-1.56&1.28&0.00&97.67&-&0.000&0.000&Trigonal&148&1.51&2\\ 
mp-510409&TiSiNi&8.35&13.78&0.00&67.57&176.81&0.014&0.000&Orthorhombic&62&1.37&1\\ 
mp-540671&Ti$_7$Cl$_1$$_6$&4.18&1.26&0.00&95.00&-&0.003&0.000&Orthorhombic&58&1.35&1\\ 
mp-540672&Ti$_7$Br$_1$$_6$&3.44&3.48&0.00&39.54&-&0.004&0.000&Orthorhombic&58&1.35&1\\ 
mp-540767&Fe$_2$(SO$_4$)$_3$&0.20&0.35&0.00&6.02&-&0.002&0.000&Trigonal&148&1.15&1\\ 
mp-540828&FeClO&0.82&0.15&0.00&0.00&42.64&0.000&0.000&Orthorhombic&59&1.15&2\\ 
mp-542028&Ti$_3$Al$_2$NiN&7.53&7.90&0.00&77.43&-&0.001&0.000&Cubic&227&1.55&1\\ 
mp-542915&TiAl$_3$&7.79&3.34&0.00&69.24&107.53&0.000&0.000&Tetragonal&139&1.98&2\\ 
mp-554429&Mn$_3$Si(O$_2$F)$_2$&2.29&0.90&0.00&91.99&-&0.000&0.000&Orthorhombic&62&1.87&2\\ 
mp-555343&Ni$_5$Se$_4$(BrO$_6$)$_2$&4.74&15.57&0.00&47.10&-&0.011&0.000&Triclinic&2&1.83&1\\ 
mp-555908&CoF$_2$&0.53&38.85&0.00&76.00&101.07&0.000&0.000&Tetragonal&136&1.57&2\\ 
mp-556560&MnF$_3$&-1.65&0.00&0.00&100.00&-&0.009&0.000&Monoclinic&15&1.68&1\\ 
mp-556911&FeF$_2$&0.97&0.07&0.00&0.00&97.77&0.012&0.000&Tetragonal&136&1.22&2\\ 
mp-558502&FeTeO$_3$F&2.63&35.81&0.00&52.17&-&0.000&0.000&Monoclinic&14&1.90&2\\ 
mp-560023&NiSnF$_6$&1.51&19.38&0.00&70.05&33.26&0.000&0.000&Trigonal&148&1.43&2\\ 
mp-560902&MnF$_2$&0.60&0.00&0.00&100.00&86.68&0.000&0.000&Tetragonal&136&1.75&2\\ 
mp-561319&CoSnF$_6$&-0.07&27.24&0.00&76.67&-&0.000&0.000&Trigonal&148&1.51&2\\ 
\bottomrule\end{tabular}}

\clearpage
\resizebox{\textwidth}{!}{\begin{tabular}{c||c||c|c|c|c|c|c|c|c|c|c|c}\toprule\multicolumn{1}{l||}{\textbf{ID}} & \multicolumn{1}{l||}{\textbf{Formula}} & \multicolumn{1}{l|}{\textbf{E$_{F}$}} & \multicolumn{1}{l|}{\textbf{Price}} & \multicolumn{1}{l|}{\textbf{E$_\text{hull}$}} & \multicolumn{1}{l|}{\textbf{NIR}} & \multicolumn{1}{l|}{\textbf{K$_\text{voigt}$}} & \multicolumn{1}{l|}{\textbf{E$_{g, \text{D}}$}} & \multicolumn{1}{l|}{\textbf{E$_{g, \text{ID}}$}} & \multicolumn{1}{l|}{\textbf{Symmetry}} & \multicolumn{1}{l|}{\textbf{SG}} & \multicolumn{1}{l|}{\textbf{Env}} & \multicolumn{1}{l}{\textbf{C}} \\ \midrule 
mp-561571&CoSn$_2$H$_1$$_2$(OF)$_6$&0.13&23.87&0.00&76.80&-&0.000&0.000&Triclinic&2&1.53&2\\
mp-567412&Y$_5$Sn$_3$&5.04&39.53&0.00&89.77&63.27&0.000&0.000&Hexagonal&193&1.59&2\\ 
mp-567678&Y$_7$FeI$_1$$_2$&4.12&40.53&0.00&62.87&-&0.000&0.000&Trigonal&148&1.15&2\\ 
mp-567705&TiAl$_2$&7.44&3.35&0.00&73.27&108.56&0.007&0.000&Tetragonal&141&1.88&1\\
mp-568056&LaCo$_9$Si$_4$&7.97&44.69&0.00&74.92&-&0.005&0.000&Tetragonal&140&1.80&2\\ 
mp-568934&Ti$_4$AlN$_3$&8.14&2.84&0.00&89.94&-&0.007&0.000&Hexagonal&194&1.47&1\\ 
mp-569196&YNi$_3$&6.30&31.04&0.00&70.76&131.96&0.000&0.000&Trigonal&166&1.32&2\\ 
mp-569610&CoI$_2$&4.08&45.32&0.00&54.90&9.31&0.000&0.000&Trigonal&164&1.73&2\\ 
mp-571143&TiCl$_3$&2.09&1.06&0.00&95.00&8.98&0.000&0.000&Hexagonal&193&1.35&2\\ 
mp-574339&Y$_2$Ni$_7$&6.40&30.44&0.00&69.29&-&0.009&0.000&Trigonal&166&1.32&2\\ 
mp-582055&La$_7$Ni$_1$$_6$&8.88&12.79&0.00&75.84&-&0.000&0.000&Tetragonal&121&1.82&2\\ 
mp-601820&Fe$_3$Co&5.32&16.71&0.00&19.78&186.99&0.000&0.000&Tetragonal&123&1.30&2\\ 
mp-601842&Fe$_9$Co$_7$&5.51&28.87&0.00&34.26&190.37&0.000&0.000&Cubic&221&1.41&2\\ 
mp-605437&FeHO$_2$&1.43&0.47&0.00&0.00&-&0.002&0.000&Orthorhombic&62&1.15&2\\ 
mp-616559&Ti$_2$S&8.24&2.58&0.00&74.42&147.46&0.001&0.000&Orthorhombic&58&1.35&1\\ 
mp-620032&Ti$_2$Se&7.40&11.81&0.00&74.66&-&0.002&0.000&Orthorhombic&58&1.78&1\\ 
mp-637255&Ti$_2$Sn$_3$&8.28&28.46&0.00&80.81&-&0.012&0.000&Orthorhombic&64&1.54&1\\ 
mp-643547&CoH$_2$SO$_5$&1.62&22.26&0.00&53.80&-&0.000&0.000&Monoclinic&15&1.73&2\\ 
mp-653429&Mn$_7$FeCl$_3$O$_1$$_0$&1.86&0.15&0.00&87.32&-&0.002&0.000&Cubic&225&1.94&2\\ 
mp-669720&Ni$_3$Sn$_2$&9.44&30.86&0.00&68.06&-&0.017&0.000&Orthorhombic&62&1.47&1\\ 
mp-672373&La$_3$Co$_4$Sn$_1$$_3$&8.91&31.79&0.00&80.31&-&0.000&0.000&Cubic&223&1.74&2\\ 
mp-672677&Ti$_6$Co$_1$$_6$Si$_7$&7.53&44.02&0.00&75.55&-&0.000&0.000&Cubic&225&1.62&2\\ 
mp-865235&Ti$_6$Al$_1$$_6$Ni$_7$&8.18&11.22&0.00&65.15&-&0.001&0.000&Cubic&225&1.72&1\\ 
mp-976187&Pr(Al$_2$Fe)$_4$&7.11&26.29&0.00&43.17&-&0.000&0.000&Tetragonal&139&1.83&2\\ 
mp-976378&La$_3$Sn$_4$&7.71&19.21&0.00&85.41&-&0.000&0.000&Orthorhombic&63&1.92&2\\ 
mp-985707&CoSn$_3$&8.59&39.28&0.00&76.86&-&0.007&0.000&Tetragonal&142&1.61&2\\ 
mp-1009077&FeH&5.93&0.39&0.00&0.00&176.89&0.000&0.000&Cubic&225&1.15&2\\ 
mp-1009133&MnCo&6.18&33.07&0.00&87.58&-&0.004&0.000&Cubic&221&1.89&1\\ 
mp-1018028&TiS&8.30&2.10&0.00&62.10&142.40&0.000&0.000&Hexagonal&187&1.35&2\\ 
mp-1025263&Ti$_2$NiS$_4$&6.73&6.41&0.00&49.70&-&0.009&0.000&Monoclinic&12&1.34&1\\ 
mp-1069204&NdCoSi$_3$&7.69&44.30&0.00&76.45&-&0.000&0.000&Tetragonal&107&1.95&2\\ 
mp-1070857&NdSiNi$_4$&7.05&35.34&0.00&69.06&-&0.006&0.000&Orthorhombic&65&1.68&1\\ 
mp-1071163&TiO&8.40&2.70&0.00&95.00&-&0.004&0.000&Hexagonal&189&1.35&1\\ 
mp-1071627&LaSiNi$_4$&8.89&15.51&0.00&68.71&-&0.000&0.000&Orthorhombic&65&1.67&2\\ 
mp-1072437&YSiNi$_4$&7.04&28.19&0.00&66.24&-&0.000&0.000&Orthorhombic&65&1.34&2\\ 
mp-1077503&TiSi$_2$&9.53&5.88&0.00&68.00&-&0.010&0.000&Orthorhombic&63&1.43&1\\ 
mp-1078634&LaFeO$_3$&5.52&0.72&0.00&67.76&-&0.000&0.000&Trigonal&167&1.98&2\\ 
mp-1078717&La(NiSn)$_2$&8.64&23.15&0.00&77.07&-&0.000&0.000&Tetragonal&129&1.73&2\\ 
mp-1078782&YCoSi$_2$&6.70&39.40&0.00&77.92&-&0.000&0.000&Orthorhombic&63&1.62&1\\ 
mp-1079037&LaSi$_2$Ni&8.94&8.10&0.00&74.91&-&0.000&0.000&Orthorhombic&63&1.90&2\\ 
mp-1079273&La$_3$Sn$_7$&8.35&23.78&0.00&83.01&-&0.000&0.000&Orthorhombic&65&1.83&2\\ 
mp-1079460&Ti$_3$Sn&7.56&17.79&0.00&86.85&-&0.007&0.000&Orthorhombic&63&1.46&1\\ 
mp-1080643&LaFeSi$_2$&8.44&2.37&0.00&62.67&-&0.000&0.000&Orthorhombic&63&1.87&2\\ 
mp-1084757&LaNiSn$_2$&8.73&22.90&0.00&79.91&-&0.000&0.000&Orthorhombic&63&1.78&2\\ 
mp-1105598&Y$_3$Co&3.43&46.78&0.00&95.66&-&0.005&0.000&Orthorhombic&62&1.73&1\\ 
mp-1105633&Y$_3$Ni&3.73&39.75&0.00&92.06&-&0.000&0.000&Orthorhombic&62&1.32&2\\ 
mp-1106074&Y$_6$CoI$_1$$_0$&3.80&42.30&0.00&65.15&-&0.000&0.000&Triclinic&2&1.73&2\\ 
mp-1188435&Y$_2$Co$_3$Si$_5$&7.35&40.54&0.00&75.83&-&0.000&0.000&Monoclinic&15&1.63&2\\ 
mp-1189066&Y$_5$Si$_3$H&4.85&37.38&0.00&91.24&-&0.004&0.000&Hexagonal&193&1.50&2\\ 
mp-1190442&FeSeO$_3$F&0.74&8.78&0.00&29.29&-&0.001&0.000&Monoclinic&14&1.77&2\\ 
mp-1190679&PrSi$_2$Ni$_9$&7.86&38.84&0.00&62.72&-&0.000&0.000&Tetragonal&141&1.49&2\\ 
mp-1190732&LaSi$_2$Ni$_9$&8.85&19.07&0.00&62.64&-&0.000&0.000&Tetragonal&141&1.52&2\\ 
mp-1191331&Ti$_2$Co&6.00&26.45&0.00&87.76&-&0.001&0.000&Cubic&227&1.50&2\\ 
mp-1192030&CoH$_2$Cl$_2$O&0.73&25.75&0.00&76.00&-&0.000&0.000&Orthorhombic&62&1.73&2\\ 
mp-1193093&LaNi$_5$Sn&9.07&21.15&0.00&70.35&-&0.006&0.000&Hexagonal&194&1.63&1\\ 
mp-1193428&PrNi$_5$Sn&7.58&47.06&0.00&70.44&-&0.000&0.000&Hexagonal&194&1.58&2\\ 
mp-1193657&Ti$_4$Co$_4$Si$_7$&8.73&27.71&0.00&72.06&-&0.002&0.000&Tetragonal&139&1.54&1\\ 
mp-1194971&Ni$_5$Se$_4$(ClO$_6$)$_2$&1.66&16.51&0.00&52.89&-&0.005&0.000&Triclinic&2&1.83&1\\ 
mp-1197725&Y$_2$Fe$_3$Si$_5$&7.46&18.09&0.00&49.61&-&0.002&0.000&Tetragonal&128&1.31&1\\ 
mp-1198388&Y$_2$Mn$_3$Si$_5$&7.32&18.16&0.00&84.01&-&0.006&0.000&Tetragonal&128&1.80&2\\ 
mp-1198642&Y$_7$Co$_6$Sn$_2$$_3$&7.80&39.25&0.00&80.77&-&0.000&0.000&Trigonal&164&1.60&2\\ 
mp-1198940&Nd$_3$Co$_4$Sn$_1$$_3$&7.97&42.62&0.00&80.42&-&0.000&0.000&Cubic&223&1.74&2\\ 
mp-1199133&Y$_1$$_1$Sn$_1$$_0$&5.87&38.72&0.00&87.39&-&0.001&0.000&Tetragonal&139&1.59&2\\ 
mp-1205373&Y$_6$CoBr$_1$$_0$&2.76&21.20&0.00&55.91&-&0.003&0.000&Triclinic&2&1.73&2\\ 
\bottomrule\end{tabular}}

\clearpage
\captionof{table}{List of parameters for a subset of widely known topological materials (including topological insulators, Dirac and Weyl semimetals, and nodal-line semimetals). E$_F$: Fermi level in electronvolts (eV). Price is in USD (in 2023) per kilogram (\$/kg). E$_\text{hull}$: energy above the complex hull; NIR: net import resilience; K$_{\text{voigt}}$: Bulk modulus (Voigt average, GPa); E$_{\text{g, D}}$: direct band gap in electronvolts; E$_{\text{g, ID}}$: indirect band gap in electronvolts; SG: space group number index; Env: environmental score. The materials are categorized (C) into three groups, with spin-orbit coupling taken into account: (1) topological insulator, (2) semimetal, and (3) trivial.}
\label{tab2}
\resizebox{\textwidth}{!}{\begin{tabular}{c||c||c|c|c|c|c|c|c|c|c|c|c}\toprule\multicolumn{1}{l||}{\textbf{ID}} & \multicolumn{1}{l||}{\textbf{Formula}} & \multicolumn{1}{l|}{\textbf{E$_{F}$}} & \multicolumn{1}{l|}{\textbf{Price}} & \multicolumn{1}{l|}{\textbf{E$_\text{hull}$}} & \multicolumn{1}{l|}{\textbf{NIR}} & \multicolumn{1}{l|}{\textbf{K$_\text{voigt}$}} & \multicolumn{1}{l|}{\textbf{E$_{g, \text{D}}$}} & \multicolumn{1}{l|}{\textbf{E$_{g, \text{ID}}$}} & \multicolumn{1}{l|}{\textbf{Symmetry}} & \multicolumn{1}{l|}{\textbf{SG}} & \multicolumn{1}{l|}{\textbf{Env}} & \multicolumn{1}{l}{\textbf{C}} \\ \midrule
mp-605&ZrTe$_5$&4.55&64.87&0.00&71.87&19.22&0.077&0.077&Orthorhombic&63&2.31&1\\ 
mp-1168&HfTe$_5$&4.64&4208.22&0.00&75.00&19.84&0.083&0.083&Orthorhombic&63&2.31&1\\ 
mp-1372&Cd$_3$As$_2$&1.87&3.51&0.00&48.07&-&0.025&0.011&Tetragonal&137&3.00&2\\ 
mp-2730&HgTe&2.21&-&0.00&75.00&36.43&0.000&0.000&Cubic&216&2.73&2\\ 
mp-3938&ZrSiS&7.60&18.99&0.00&41.23&125.62&0.026&0.000&Tetragonal&129&1.50&1\\ 
mp-4628&ZrSiSe&6.55&23.24&0.00&49.29&101.67&0.018&0.000&Tetragonal&129&2.10&1\\ 
mp-9830&NbP&7.24&18.67&0.60&100.00&167.29&0.015&0.000&Tetragonal&141&1.79&1\\ 
mp-12597&TaP&7.76&128.47&0.69&100.00&-&0.013&0.000&Tetragonal&141&2.36&1\\ 
mp-19807&Co$_3$(SnS)$_2$&8.13&41.11&0.00&68.05&97.08&0.000&0.000&Trigonal&166&1.65&2\\ 
mp-21171&MnGaCo$_2$&5.49&215.05&0.00&88.34&189.71&0.004&0.000&Cubic&225&1.83&1\\ 
mp-22389&Mn$_3$Sn&6.93&14.74&0.18&90.37&90.03&0.000&0.000&Hexagonal&194&1.86&2\\ 
mp-27838&Na$_3$Bi&1.36&7.04&0.00&96.00&17.16&0.001&0.001&Hexagonal&194&2.64&2\\ 
mp-34202&Bi$_2$Te$_3$&5.17&37.95&0.00&85.96&19.93&0.346&0.249&Trigonal&166&2.48&1\\ 
mp-541837&Bi$_2$Se$_3$&3.69&13.44&0.00&79.36&19.90&0.410&0.397&Trigonal&166&2.52&1\\ 
mp-1070314&Eu(CdAs)$_2$&3.54&11.19&0.00&66.54&-&0.000&0.000&Trigonal&164&2.55&2\\ 
\bottomrule\end{tabular}}

\twocolumn

\section*{Methods}
\subsection*{Topological materials database}
The list of topological materials in this database with a viewpoint on sustainability currently includes 16,858 topological materials, with 4,768 entries for topological insulators and 11,980 entries for topological semimetals after down-selection based on criteria set forth in the Methods. The database is derived from the original topological materials database \cite{Database} and Materials Project \cite{jain2013commentary}.

\subsection*{Thermodynamic stability, synthesis ability, and CMOS compatibility}
One commonly used material property to predict stability is the energy above the convex hull \cite{bartel2022review}. The convex hull is formed by plotting the formation energy of competing phases with respect to composition. It connects phases that have lower energy than any other phase or a linear combination of phases at that composition. Phases lying on or within the convex hull are thermodynamically stable, while those above the convex hull are unstable. To determine the energy above the convex hull for topological materials, data is extracted from the Material Project database \cite{jain2013commentary} using the Materials Project ID through an API. The convex hull of materials in the Material Project database is constructed from a first-principle calculation of formation energies with a r2SCAN metaGGA functional. Note that this formation energy is calculated at 0 K. Therefore, the energy above the convex hull used in this analysis only implies the stability of the materials at zero temperature. 

\subsection*{Price and supply chain of raw materials }
The price per kilogram of each topological material is calculated by taking the weighted average of the elements present in the compound. For example, the price of $A_xB_y$ is calculated from 
\[ P_{A_xB_y} = xP_AM_A + yP_BM_B / xM_A + yM_B. \]

The stability of the supply chain is complex as the raw materials have to pass through many processes of extraction, separation, and refinement. These processes are not necessarily done in the same countries depending on the technology requirement and sometime require cooperation with other materials in the process. Therefore, the disruption in any step of the production of the material can affect the supply chain of not only its own materials but different materials as well. For simplicity, this analysis uses net import resilience (NIR) to represent the supply chain stability in the United States \cite{ScienceBase_2017}. For raw non-critical elements without NIR data, the value of NIR is decided based on the natural abundance of the elements. Gas and common elements, including O, N, F, Cl, Br, I, Na, and K, are assumed to be 0 and 100 otherwise. For each material listed in the database, the NIR score for the materials is computed using a similar method to the price calculation, employing the weighted average of the elements based on their composition. 

\subsection*{Environmental impact of raw materials}
The assessment of the environmental effects of raw materials involves four indices: the environmental hazard potential, specific cumulative mass, energy demand, and the environmental government index \cite{EnvCri}. The ÖkoRess project report provides an analysis of 61 raw materials. The EHP score assesses the production process using eight criteria, including
\begin{enumerate}
  \item Environmental damage resulting from geochemical conditions in mining
  \item Concentration of heavy metals at the mining site
  \item The presence of radioactive substances (uranium and thorium) at the mining site
  \item Mine type (underground, open pit, alluvial)
  \item Use of auxiliary chemicals in extraction and processing
  \item Risk of natural disasters
  \item Water stress index
  \item Location of the mine in protected areas or the Arctic region
\end{enumerate}

The second index is the specific cumulative mass and energy demand. The specific cumulative mass demand (CRD$_{\text{specific}}$) represents the mass required from the mining process to produce one ton of material. Similarly, the specific cumulative energy demand (CED$_{\text{specific}}$) represents the energy (in MJ) needed to produce one ton of material. In the ÖkoRess project report, these values are converted into global cumulative mass and energy flows by multiplying them with the total annual production of the materials. This provides the total mass extraction and energy consumption in material production per year. The global cumulative mass and energy demand are then categorized into three levels, from low to high, based on the distribution among all the raw materials studied (low: less than 25\% quartile, high: more than 75\% quartile, and medium otherwise). However, in this analysis, the specific cumulative demands are used directly without converting to global cumulative per year. The raw materials are then classified into three categories, from low to high, in a similar manner.

The third index assesses the governmental perspective. From the ÖkoRess project report, this score is calculated using the environmental performance index (EPI), which rates countries on a scale of 0 to 100 based on their performance in various categories related to environmental issues. For each raw material, the EGov score is calculated from the EPI score of the mining country weighted by the share of the country's material production. 
\[ \text{EGov}_{\text{raw material}} = \sum_{\text{country}} \text{EPI}_{\text{country}}\cdot \text{X}_{\text{country}}\]
where X is the percent production share of the country. This index reflects the societal impact of raw material production on a global scale. The raw materials are then categorized into three categories (high: less than 25\% quartile, low: more than 75\% quartile, and medium otherwise) with a number from 1 to 3 assigned. 

Lastly, the recyclability (R) of the material is considered. The recyclability fraction indicates the proportion of materials that can be recovered after their end-of-life or use. The European Union Commission has reported the recyclability of specific materials \cite{doi/10.2873/167813}. It should be noted that the recyclability of topological materials may not align with existing recycling and separation techniques since it can only be determined during the product design stage. Nevertheless, utilizing materials with high recyclability can help reduce the environmental impact of mining and extraction processes.

To reduce the dimension of the indices for facilitating the screening process, these environment scores are combined in the following way
\begin{equation}
\resizebox{0.45\textwidth}{!}{
  $\text{Env}_{\text{A}_x\text{B}_y\cdots} = \frac{\sum_{(\text{AB}\cdots)} (\text{EHP}_{A}+\text{CED}_{A}+\text{EGov}_{A})(100-\text{R})xM_A}{\sum_{(\text{AB}\cdots)}x\text{M}_A}$
}.
\end{equation}

This Env score signifies the environmental (EHP, CED) and social (EGov) impact of the production of raw materials for synthesizing topological materials. For materials that consist of elements that are not reported in the study, the contribution of these elements to the Env score is neglected (by considering a weight of 0 for non-critical elements). In addition, we assign the Env score of 3 to Cd, Pb, Hg, As, ans Sb due to their toxicity and 5 to radioactive elements, including Pu, AM, Cm, Bk, Cf, Es, Fm, MD, No, Lr, Rf, Db, Sg, Bh, Hs, Mt, Ds, Rg, Cn, NH, U, FL, Lv, Ts, Og, Tc, Po, Pa and Os due to its rarity. 

\section*{Data availability}

\section*{Code availability}

\bibliography{sn-bibliography}

\section*{Acknowledgements}
TB and ML acknowledge the support from National Science Foundation (NSF) Convergence Accelerator Award No. 2235945. NCD acknowledges the support by US Department of Energy (DOE), Office of Science (SC), Basic Energy Sciences (BES), Award No. DE-SC0020148. TN is supported by NSF Designing Materials to Revolutionize and Engineer our Future (DMREF) Program with Award No. DMR-2118448, and MathWorks Fellowship. The authors acknowledge the helpful discussions with L Molnar, J Depew, BA Bernevig, A Bansil, J Cha, CZ Chang, J Checkelsky, B Chen, CT Chen, L Fu, R Han, S Honeyman, I Garate, L Liu, R Movassagh, T Momiki, KE Nelson, J Provine, N Regnault, M Soljacic, O Viyuela, and S Xu. ML acknowledges the support from the MIT Class of 1947 Career Development Professor Chair, the MIT-IBM Program, and the support from R Wachnik.

\section*{Competing interests}
The authors declare no competing interests.

\setcounter{table}{0}
\renewcommand{\thetable}{S\arabic{table}}
\include{sn-supplementary}

\end{document}